\documentclass[twocolumn,showpacs,preprintnumbers,amsmath,amssymb]{revtex4}

\usepackage{graphicx}
\usepackage{dcolumn}
\usepackage{bm}

\begin{document}


\title{Quantum glassiness in clean strongly correlated systems: an example of
  \\ topological overprotection}

\author{Claudio Chamon}

\affiliation{
Physics Department, Boston University,
Boston, MA 02215, USA
}


\begin{abstract}
  
  This paper presents solvable examples of quantum many-body Hamiltonians of
  systems that are unable to reach their ground states as the environment
  temperature is lowered to absolute zero. These examples, three dimensional
  generalizations of quantum Hamiltonians proposed for topological quantum
  computing, 1) have no quenched disorder, 2) have solely local interactions,
  3) have an exactly solvable spectrum, 4) have topologically ordered
  ground states, and 5) have slow dynamical relaxation rates akin to those of
  strong structural glasses.

\end{abstract}


\maketitle

Describing matter at near absolute zero temperature requires understanding a
system's quantum ground state and the low energy excitations around it, the
quasiparticles, which are thermally populated by the system's contact to a
heat bath. However, this paradigm breaks down if thermal equilibration is
obstructed. While such non-equilibrium behavior may be expected in disordered
and frustrated quantum systems (like for instance quantum spin
glasses~\cite{Bray-Moore}, long-range Josephson junction arrays in a
frustrating magnetic field~\cite{Kagan-Feigelman-Ioffe1999}, or
self-generated mean-field glasses~\cite{Westfahl-etal2003}), it is
non-obvious that it may exist in clean systems with only local interactions
and without a complicated distribution of energy levels. In this paper I
present solvable examples, three dimensional generalizations of Hamiltonians
proposed for topological quantum computing, that have solely local
interactions, no quenched disorder, and relaxation rates akin to those of
strong structural glasses. Therefore, in these systems the topologically
ordered ground states are not reached when the temperature is reduced to
absolute zero.

Topological order and quantum number fractionalization are some of the most
remarkable properties of systems of strongly interacting particles. Some
phases of matter, in contrast to common examples like crystals and magnets,
are not characterized by a local order parameter and broken symmetries.
Instead, as shown by Wen~\cite{Wen1990,Wen1995}, some quantum phases are
characterized by their topological order, such as the degeneracy of the
ground state when the system is defined on a torus or other surface of higher
genus. These topological degeneracies cannot be lifted by {\it any} local
perturbation. Topological order and quantum number fractionalization are
intimately related, and much effort has recently been directed at these
exotic properties, for they may play a role in the mechanism for
high-temperature
superconductivity~\cite{Anderson1987,Laughlin1988,Senthil-Fisher2001}. Also,
the robustness of a topological degeneracy to local noise due to an
environment is at the core of the idea behind topological quantum
computation, as proposed by Kitaev~\cite{Kitaev97}. Interestingly enough,
strong correlations that can lead to these exotic quantum spectral properties
can in some instances also impose kinetic constraints, similar to those
studied in the context of classical glass
formers~\cite{Ritort&Sollich03,FreAnd84,FreAnd85,Garrahan&ChandlerPNAS,Alvarez-Franz-Ritort96,Lipowski97,Buhot&Garrahan02,Garrahan02}.

The possibility of glassiness in pure strongly correlated quantum systems
with solely local interactions is demonstrated by studying the following
exactly solvable example. 
A model displaying strong like glassiness is constructed on a
three-dimensional (3D) face-centered cubic (fcc) Bravais lattice, spanned by
the primitive vectors $\bm
a_1=\left(\frac{1}{\sqrt{2}},\frac{1}{\sqrt{2}},0\right)$, $\bm
a_2=\left(0,\frac{1}{\sqrt{2}},\frac{1}{\sqrt{2}}\right)$, and $\bm
a_3=\left(\frac{1}{\sqrt{2}},0,\frac{1}{\sqrt{2}}\right)$.
Each site can be indexed by $i,j,k\in \mathbb Z$, denoted by a superindex
$I\equiv (i,j,k)$. At every lattice site $I$ one defines quantum spin $S=1/2$
operators $\sigma^{\rm x}_{I}$, $\sigma^{\rm y}_{I}$, and $\sigma^{\rm
  z}_{I}$.

\begin{figure}
\vspace{0cm}
\includegraphics[angle=0,scale=0.27]{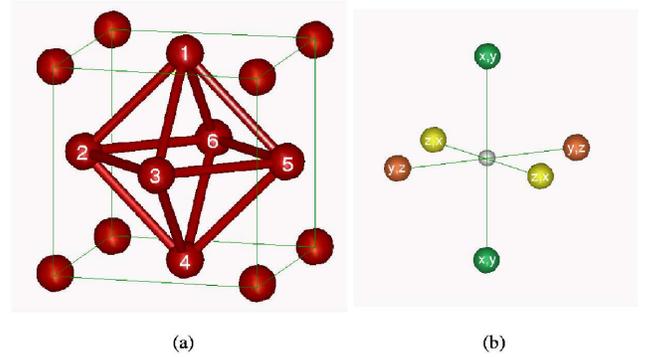}
\caption{
\label{fig:fcc} 
(a) Cubic cell of an fcc lattice. The centers of the six faces form an
octahedron, with its sites labeled from 1 (topmost) to 6. In addition to the
set of octahedra formed by the face centered sites, there are three more sets
of octahedra that can be assembled from sites both on faces and on corners of
the cubic cells, totaling 4 such sets. Six-spin operators are defined on
these octahedra using the $\sigma^{\rm x,y,z}$ components of spin on each
vertex as described in the text.\\
(b) Centers of 6 octahedra cells that share a spin, which resides at the site
$\tilde I$ shown at the center. The x,y, or z labels sitting at the centers
of the octahedra show which spin operator $\sigma^{\rm x,y,z}_{\tilde I}$
flip their $O_I$ eigenvalue. Acting with any of the operators $\sigma^{\rm
  x,y,z}_{\tilde I}$ always flip the eigenvalues $O_I$ of exactly four
octahedra.
}
\end{figure}

The fcc lattice can house sets of {\it octahedra}: the simplest one to
visualize is the one assembled from the centers of the six faces of a cubic
cell, and is shown in Fig.~\ref{fig:fcc}(a). In addition to this simple
set, there are three more sets of octahedra that can be assembled from sites
both on faces and on corners of the cubic cells, totaling 4 such sets, which
we label by $A,B,C$ and $D$.

It is simple to see that the total number of octahedra equals the number of
spins: each lattice site $I$ is the topmost vertex of a single octahedron.
Define $P_I$ as the set of six lattice points forming the octahedron with
site $I$ at its top. The six vertices are indexed by $J_n(I)$, for
$n=1,\dots, 6$, with one of the vertices $J_1(I)=I$. The six labels are
assigned in such as way that the pairs $\{J_1,J_4\},\{J_2,J_5\},\{J_3,J_6\}$
are diagonally opposite sites from one another, and this number labeling is
illustrated for a single octahedron in Fig.~\ref{fig:fcc}(a). From the
one-to-one relation between a site $I$ and an octahedron $P_I$, we can also
partition the lattice sites into four sublattices $A,B,C$ and $D$.

Now define the operators ${\cal O}_{I}$ as
\begin{equation}
{\cal O}_{I}=
\sigma^{\rm z}_{J_1(I)}\;
\sigma^{\rm x}_{J_2(I)}\;
\sigma^{\rm y}_{J_3(I)}\;
\sigma^{\rm z}_{J_4(I)}\;
\sigma^{\rm x}_{J_5(I)}\;
\sigma^{\rm y}_{J_6(I)}
\;.
\label{eq:O-octa-def}  
\end{equation}
This construction generalizes to 3D the plaquette interactions defined for
planar 2D lattices by Kitaev~\cite{Kitaev97} and Wen~\cite{Wen2003c}. These
operators commute, $[{\cal O}_I,{\cal O}_{I'}]=0$ for all pairs $I,I'$.  It
is simple to see how: two octahedra $P_I$ and $P_{I'}$ can either share 0,1,
or, at most, 2 spins. If they share 0 spins, they trivially commute. If they
share 1 spin, the component (${\rm x,y}$ or $\rm z$) of $\bm \sigma$ for that
shared spin coincides for both ${\cal O}_I$ and ${\cal O}_{I'}$ (the two
octahedra touch along one of their diagonals). If they share 2 spins, the
components of $\bm \sigma$ used in the definition of ${\cal O}_I$ and ${\cal
  O}_{I'}$ are different for both spins, there is a minus sign from commuting
the spin operators from each of the shared spins, and the two minus signs
cancel each other.

Consider the Hamiltonian
\begin{equation}
\hat H=-\frac{h}{2}\;\sum_I {\cal O}_I
\;.
\label{eq:Hamiltonian}  
\end{equation}
Because the ${\cal O}_I$ all commute, the eigenvalues of the Hamiltonian can
be labeled by the list of eigenvalues $\{O_I\}$ of all the ${\cal O}_I$.
Notice that ${\cal O}_I^2=\openone$, and so each $O_I=\pm 1$. In particular,
the ground state corresponds to $O_I=1$ for all $I$.

Because the number of spins equals the number $N$ of sites and of octahedra,
one may naively expect that the list $\{O_I=\pm 1\}$ exhausts the $2^N$
states in the Hilbert space. However, there are constraints that the ${\cal
  O}_I$ satisfy when the system is subject to periodic boundary conditions
(compactified). One can show that
\begin{equation}
\prod_{I\in A} {\cal O}_I=
\prod_{I\in B} {\cal O}_I=
\prod_{I\in C} {\cal O}_I=
\prod_{I\in D} {\cal O}_I=
\openone
\;.
\label{eq:constraints-3D-fcc}  
\end{equation}
There are four constraints; therefore there are only $2^{N-4}$ independent
$\{O_I=\pm 1\}$. This implies, in particular, that there is a ground state
degeneracy of $2^4=16$. The ground state degeneracy is {\it not}
associated with a symmetry; in particular, it is easy to show that $\langle
\sigma^{\rm x,y,z}_I\rangle=0$. This is a topological degeneracy, and the
eigenvalues $T_a=\pm 1$ ($a=1,2,3,4$) of a set of four non-local
(topological) operators ${\cal T}_{a}$ are needed to distinguish between the
16 degenerate ground states.

The operators ${\cal T}_{a}$ can be constructed as follows. Let ${\cal
  P}_l=\{I|j+k=l\}$ be a set of points along a horizontal plane. Notice that
  each such plane contains sites in only two of the four sublattices
  $A,B,C,D$. For example ${\cal P}_{1}\subset A\cup B$ and ${\cal
  P}_{2}\subset C\cup D$.  Define
${\cal T}_{1}=\prod_{I\in {\cal P}_{1}\cap A}\sigma^{\rm z}_I,\;
{\cal T}_{2}=\prod_{I\in {\cal P}_{1}\cap B}\sigma^{\rm z}_I,\;
{\cal T}_{3}=\prod_{I\in {\cal P}_{2}\cap C}\sigma^{\rm z}_I,\;
{\cal T}_{4}=\prod_{I\in {\cal P}_{2}\cap D}\sigma^{\rm z}_I\;
.$
It is simple to check that $[{\cal T}_{a},{\cal O}_I]=0$ for all $a$ and $I$,
and the ${\cal T}_{a}$ trivially commute among themselves. Hence the
four eigenvalues ${T}_{1,2,3,4}=\pm 1$ of ${\cal T}_{1,2,3,4}$ can
distinguish the 16 degenerate ground states.

The spectrum of the Hamiltonian Eq.~(\ref{eq:Hamiltonian}) is that of a
trivial set of $N-4$ free spins, determined by the list of eigenvalues
$\{O_I=\pm 1\}$ of all the ${\cal O}_I$, subject to the condition
Eq.~(\ref{eq:constraints-3D-fcc}): $E_{\{O_I\}}=-\frac{h}{2}\sum_I O_I$.
Excitations above the ground state ($O_I=1$ for all $I$) are ``defects''
where $O_I=-1$ at certain sites $I$. The equilibrium partition function is
given by $Z=16\sum_{\{O_I=\pm1\}} e^{\;\frac{1}{2}\beta h\sum_I O_I}$. At thermal
equilibrium at temperature $T$, the thermal average $\langle O_I\rangle =
\tanh \frac{h}{2T}$, and the concentration or density of $O_I=-1$ defects is
$c=\frac{1}{2}\left(1-\tanh \frac{h}{2T}\right)$. Notice that we have
encountered an analogous situation to that in the classical spin facilitated
models~\cite{Ritort&Sollich03}, in particular the plaquette models displaying
glassy
dynamics~\cite{Alvarez-Franz-Ritort96,Lipowski97,Buhot&Garrahan02,Garrahan02}:
the thermodynamics is trivial in terms of non-interacting defect variables.
To study the approach to such an equilibrium state, the coupling of the
system to a bath of quantum oscillators must be introduced.

The Hamiltonian of the system plus bath of oscillators can be formulated as
~\cite{Feynman-Vernon63,
Caldeira-Leggett83}
$$\hat {\cal H} = \hat H+ \hat H_{\rm bath}+\hat H_{\rm spin/bath}$$
where
$\hat H$ is defined in Eq.~(\ref{eq:Hamiltonian}), the bath $H_{\rm bath}$
contains a family of harmonic oscillators ${\bm a}_{\lambda,I},{\bm
  a}^\dagger_{\lambda,I}$ for each site, and
\begin{equation}
H_{\rm spin/bath}= 
\sum_{I,\alpha} g_\alpha\; \sigma^\alpha_I \; 
\sum_\lambda 
\left(
{a}_{\lambda,I}^\alpha+{{a}_{\lambda,I}^\alpha}^\dagger
\right)
\;,
\label{eq:Hamiltonian-spin+bath}
\end{equation}
where the $g_\alpha$ are the generic coupling constants for each of the three
components ($\alpha=1,2,3$ or $\rm x,\rm y,\rm z$) of the spins.

Although the spectrum of $\hat H$ in Eq.~(\ref{eq:Hamiltonian}) is the same
as that of free spins in a uniform magnetic field $h$, the variables $O_I$
for different octahedra $P_I$ {\it cannot} be independently changed, as
opposed to spin variables in a free spin model in a field $h$. The bath
couples to the {\it physical} degrees of freedom, the spins
$\bm\sigma_{\tilde I}$.  Acting on a site ${\tilde I}\in P_I$ with one of the
operators $\sigma_{\tilde I}^{\rm x},\sigma_{\tilde I}^{\rm y}$, or
$\sigma_{\tilde I}^{\rm z}$ flips or not the eigenvalue $O_I$ depending on
whether $\sigma_{\tilde I}^{\rm x,y,z}\;{\cal O}_I=\mp{\cal
  O}_I\;\sigma_{\tilde I}^{\rm x,y,z}$, respectively.  However, the spin
$\bm\sigma_{\tilde I}$ is shared by six neighboring octahedra, and thus one
cannot change the eigenvalue of $O_I$ without changing the eigenvalues
$O_{I'}$ of some of the neighbors by the action of the {\it local} spin
operator.

If integrated out, the bath degrees of freedom lead to a non-local in time
action and to dissipation effects. Instead of working with the dissipative
action, let us follow the time evolution of the system plus bath, and look at
the possible evolution pathways of the quantum mechanical amplitudes of the
system plus bath degrees of freedom. (Yet another alternative is to work
within the von~Neumann density matrix formalism~\cite{vonNeumann27}, and
follow the time evolution of the matrix elements). After evolution for time
$t$ from some initial state, the system is in a quantum mechanical
superposition
\begin{equation}
|\Psi(t)\rangle =
\!\!\!\!\!\!\!\!
\sum_{\{T_a,O_I=\pm 1\}} 
\!\!\!\!
\Gamma_{\{T_a,O_I\}}(t)\;|{\{T_a,O_I\}}\rangle
\otimes|\Upsilon_{\{T_a,O_I\}}(t)\rangle
\;,
\label{eq:state}
\end{equation}
where $|\Upsilon_{\{T_a,O_I\}}(t)\rangle$ is some state in the bath Hilbert
space with norm one. The fact that the bath degrees of freedom couple to
single quantum spins $\bm \sigma_I$ [as in
Eq.~(\ref{eq:Hamiltonian-spin+bath})] enters the problem through the
permitted channels for dynamically transferring amplitudes among the
$\Gamma_{\{T_a,O_I\}}$.

The processes that redistribute or transfer amplitude among the
$\Gamma_{\{T_a,O_I\}}$ correspond to different orders in perturbation theory
on the $g_\alpha$ system-bath coupling. There is also a thermal probability
factor coming from the bath that depends on the difference between the
initial and final energy $E_{\{O_I\}}=-\frac{h}{2}\sum_I O_I$ of the system.
The simplest class of paths is a {\it sequential} passage over states
connected through first order in $g_\alpha$ processes; this is a
``semi-classical'' type trajectory.

Through the action of a local $\sigma^{\rm x,y,z}_{\tilde I}$ operator,
exactly 4 of the 6 octahedra operators ${\cal O}_I$ sharing spin $\tilde I$
are flipped.  The reason is that the 6 octahedra operators can be divided
into 3 groups of 2 octahedra having in their definitions
Eq.~(\ref{eq:O-octa-def}), respectively, the $\rm x$,$\rm y$, and $\rm z$
component of spin operator at the shared site. This is illustrated in
Fig.~\ref{fig:fcc}(b). Acting with either of the three components of the
spin operator $\sigma^{\rm x,y,z}_{\tilde I}$ on this shared site will flip
exactly 4 out of 6 defect variables $O_{I}$. Hence, a single defect cannot be
annihilated in this process. Defects disappear from the system only through
recombination.  This multi-defect type dynamics makes it difficult for the
system to relax to equilibrium, exactly as in the kinetically constrained
classical
models~\cite{Alvarez-Franz-Ritort96,Lipowski97,Buhot&Garrahan02,Garrahan02}.
For example, if the temperature is lowered, in order to decrease the defect
density, either four defects must come together and annihilate ($4\to 0$
decay), or three defects become one ($3\to 1$ decay). Moreover, single
defects cannot simply diffuse through the system; that would require flipping
only two neighboring octahedra, but instead four are always flipped. To move,
an isolated single defect must first decay into three defects ($1\to 3$
production) because of the multi-defect dynamics, then a pair can diffuse
freely ($2\to 2$) and recombine with another defect through a $3\to 1$ decay
process. Because of the initial $1\to 3$ production process, there is an
energy barrier of $2h$ to be overcome. This activation barrier leads to
recombination/equilibration times $$t_{\rm seq.}\sim\tau_0\exp(2h/T)$$
that
grow in an Arrhenius fashion as temperature is lowered ($\tau_0$ is a
microscopic time scale).


What about quantum tunneling? Amplitude can be transfered from some initial
to some final state via virtual processes, in which the number of defects is
larger in the intermediate (virtual) steps. Virtual processes of $n$th order
involve a product of $n$ spin operators, ${\cal F}\equiv\prod_{s=1}^n
\sigma^{\alpha_s}_{{\tilde I}_s}$. For a single defect to disperse through
quantum tunneling, an ${\cal F}$ operator that flips only two octahedra is
needed. However, one can show that any ${\cal F}$ will flip {\it at least
  four} octahedra (as opposed to Kitaev's and Wen's models, in which two
defects stand at the endpoints of ``strings'', here four or more defects lie
at the corners of ``membranes''). Although defects cannot disperse, tunneling
still contributes to defect annihilation and to defect pair motion. In
perturbation theory, a process in which a defect pair separated by a distance
$\xi$ can hop by a lattice spacing has an amplitude of order $(g/h)^\xi$
(notice the energy denominator $h$) and defect annihilation has an amplitude
$(g/h)^{\xi^2}$. Hence virtual processes are suppressed exponetially in
$\xi$, and if the system were to equilibrate at temperature $T$, where the
typical defect separation is $\xi=c^{-1/3}\sim e^{h/3T}$, the characteristic
time scale $$t_{\rm tun.}\sim\tau_0 \exp\left[\ln(h/g)\;e^{h/3T}\right]$$
would grow extremely quickly as the temperature is lowered. What we learn
from this simple estimate is that quantum tunneling is less effective than
classical sequential processes in thermalizing the system. This is
counterintuitive to the notion that at low temperatures quantum tunneling
under energy barriers remains an open process while classical mechanisms are
suppressed due to high thermal activation costs. The reason for the
particular quantum freezing in this system is simple: as the distance between
defects increase at lower temperatures, the barrier {\it widths} increase,
which debilitates tunneling.  In passing, we note that in a finite system of
size $L$, one must replace $\xi$ by $L$ in the estimation of the
recombination/equilibration times, $t_{tun.}\sim\tau_0
\exp\left[\ln(h/g)\;L\right]$; this time scale is also of the order of that
for tunneling between two topological ground states in a finite system of
size $L$~\cite{Kitaev97}.


Because $t_{\rm seq.}$ and $t_{\rm tun.}$ grow rapidly as the temperature
lowers, the system will fall out of equilibrium at low temperatures, and
physical correlation functions will not be those simply computed in the
framework of equilibrium quantum statistical mechanics. The simplest
correlation function that illustrates this point is the one-point function
related to the time-dependent spatial concentration of defects
$\rho_I(t)=\frac{1}{2}[1-\langle{\cal O}_I(t)\rangle]$; let us find how it
approaches, as a function of time, the asymptotic equilibrium value $c_{\rm
  f}=\frac{1}{2}\left(1-\tanh \frac{h}{2T_{\rm f}}\right)$ when the
temperature is, say, reduced by half from $T_{\rm i}$ to $T_{\rm f}=T_{\rm
  i}/2$. The mechanism for equilibration is diffusion-annihilation of
defects. We have argued that defects are always flipped in quadruplets, and
single defects cannot freely diffuse without generating more defects. Defect
pairs, however, are free to diffuse quickly, so we can reduce the problem to
an effective reaction-diffusion~\cite{Odor-RMP2004} of the
$A+A\rightleftharpoons 0$ type for the single defects facilitated by the pair
motion. The quantum average over the state Eq.~(\ref{eq:state}) can restore
translational invariance of the density $\rho_I(t)$, so spatially homogeneous
initial densities remain homogeneous under time evolution, $\rho_I(t)=c(t)$,
hence the dynamics for this problem is controlled by the simple rate equation
${\dot c}(t)=-k (c^2-c^2_{\rm f})\;,$ with the kinetic rate coefficient $k$
directly proportional to the defect diffusion constant at temperature $T_{\rm
  f}$, from which it follows that $k\propto 1/t_{\rm seq.}$. At long times
$c(t)-c_{\rm f}\propto c_{\rm f} \;\exp(-2c_{\rm f}k\;t)$, from which we
extract the time constant for the relaxation of the one-point correlations to
be $\tau_{\rm 1pt}=t_{\rm seq.}/2c_{\rm f}$. Notice that the relaxation time
$\tau_{\rm 1pt}$ for the one-point correlation is longer than $t_{\rm seq.}$
because the annihilation rate is reduced for low densities of defects. This
enhancement must be cut off when the density of defects is of order $1/L^3$,
in which case $\tau_{\rm 1pt}\sim L^3\; t_{\rm seq.}$. The relaxation time is
just polynomial in the system size $L$, so if $t_{\rm seq.}$ saturated to a
constant value as temperature is lowered, the system would not behave as a
glass. It is the Arrhenius form of $t_{\rm seq.}$ that causes the glassy
behavior.


The fact that the system presented above is exactly solvable helps to
understand the origin of its glassy behavior, but it is {\it not} a necessary
ingredient. To illustrate this point, one can simply add a perturbation
$\delta \hat H=\sum_{I,\alpha} \Delta_\alpha\;\sigma_I^\alpha$. For
$\Delta_\alpha/h$ less than order unity, this interaction can be analyzed in
perturbation theory similarly to the arguments above for the ratios
$g_\alpha/h$. The perturbation will give the defects some mobility, but that
becomes exponentially small as the defects grow apart. Indeed, these
arguments can be generalized for any local perturbation written in terms of
the physical spins $\bm\sigma_I$, as long as the coupling constants are small
compared to the gap $h$.

To summarize, the essence of why this quantum system is glassy is the
following. The thermodynamics is best described by working in the basis of
eigenstates or defects $|\{{\cal T}_a,{\cal O}_I\}\rangle$; however, upon
acting on these states with physical spin operators $\sigma^{\rm
  x,y,z}_I$, single defects can neither be annihilated nor simply moved
around (diffused). The lack of defect diffusion in these glassy systems is
protected by the fact that any physical spin operators must flip quadruplets,
not pairs, of defect variables. The system can only relax by multi-defect
real processes that are thermally suppressed or else by virtual processes that
involve quantum tunneling of increasingly large objects as the defect density
is reduced at low temperatures.

Many elastic, thermal, electronic, and magnetic properties of classical
glassy material systems are consequences of these materials' being out of
equilibrium. Such properties can be tailored according to preparation schemes
-- for example, by controlling cooling rates. In contrast, because of the
difficulties in studying real-time dynamics of strongly interacting quantum
systems coupled to a thermal bath, very little is currently known about
properties of quantum matter that can be engineered by keeping systems out of
equilibrium. In a broader scope, the main result of this work is that it
presents a concrete example of a solvable toy model which shows without
arbitrary or questionable approximations that a pure quantum system with only
local interactions can, indeed, stay out of equilibrium. This result supports
the possibility that there may be material properties due to non-equilibrium
glassy behavior in quantum matter. It also suggests a new design constraint
for topological quantum computing: that the ground state degeneracy is
protected while the system is still able to reach the ground states.

The author thanks G.~Biroli, C.~Castelnovo, A.~Castro-Neto, L.~Cugliandolo,
M.~P.~Kennett, C.~Nayak, E.~Novais, P.~Pujol, D.~Reichman, A.~Sandvik, and
X.-G.~Wen for illuminating discussions. This work is supported in part by the
NSF Grant DMR-0305482.

\vspace{-.2in}

\bibliography{quantumglass_bib}

\end{document}